# Brilliant attosecond γ-ray emission and high-yield positron production from intense laser-irradiated Nano-Micro array

Liang-qi Zhang,[1,2] Shao-dong Wu,[1,2] Hai-rong Huang,[1,2] Hao-yang Lan,[1,2] Wei-yuan Liu,[3] Yu-chi Wu[4], Yue Yang[4], Zong-qing Zhao[4], Zhi-chao Zhu,[1,2] Wen Luo[1,2*]

[1]School of Nuclear Science and Technology, University of South China, Hengyang 421001, People's Republic of China

[2]National Exemplary Base for International Sci & Tech. Collaboration of Nuclear Energy and Nuclear Safety, University of South China, Hengyang 421001, China

[3]Key Laboratory for Laser Plasmas (MoE), School of Physics and Astronomy, Shanghai Jiao Tong University, Shanghai 200240, People's Republic of China

[4]Research Center of Laser Fusion, China Academy of Engineering Physics, Mianyang 621900, People's Republic of China

wenluo-ok@163.com

**Abstract:** We investigate a novel scheme for brilliant attosecond γ-ray emission and high-yield positron production, which is accomplished with an ultra-intense laser pulse incident upon a Nano-Micro array (NMA) with substrate incorporated. This scheme is able to realize effectively electron acceleration and colliding geometry. Both the γ-ray flash and positron bunch are then generated with high conversion efficiency. At laser intensity of $8 \times 10^{23}$ W/cm$^2$, ~27% of the laser energy is transferred successfully into the γ-rays, and ~0.7% of the laser energy into the positrons. As a consequence, ultra-short (~440 as) and ultra-brilliant (~$10^{24}$ photons s$^{-1}$ mm$^{-2}$ mrad$^{-2}$ per 0.1% BW @ 15 MeV) γ-ray burst, and high-yield ($1.48 \times 10^{11}$) and overdense (~$10^{22}$ cm$^{-3}$) positron bunch are generated. We found a sub-linear scaling of laser-to-photon conversion efficiency ($\propto I_0^{0.75}$) and a super-linear scaling of laser-to-positron conversion efficiency ($\propto I_0^{2.5}$) with the laser intensity. Multi-dimensional particle-in-cell simulations show that particle (γ photon and positron) generation can be manipulated by laser-focusing position, and NMA's length and spacing. Optimal conditions for particle generation in NMAs are obtained, indicating that microwire array has the advantage over nanowire array in particle generation in the extreme laser fields. Furthermore, positron annihilation effect in high-energy-density (HED) environment is discussed. The scheme using NMAs would provide effective avenues toward investigating attosecond nuclear science and HED physics with the coming 10 PW laser facilities.

## 1. INTRODUCTION

With the construction of multi-PW laser facilities such as Extreme Light Infrastructure (ELI)[1] and Shanghai Super-intense Ultrafast Laser Facility (SULF)[2], laser intensity above $10^{23}$ W/cm$^2$ will be accessible in the near future. In the interaction between a laser at such an intensity and matter, the electron dynamics enters into a highly nonlinear regime dominated by either radiation reaction effect or quantum electrodynamics (QED) effects.[3-6] These effects can result in violent emission of γ-rays and prolific production of positrons.[7-14] There is an increasing demand for the generation of extremely short γ-ray pulses at different photon energies with high brightness. Sufficiently high energy and bright γ-rays are useful to simulate the celestial process and extreme environments[15, 16], and to study various nuclear interactions for medical isotope production[17-19] and nuclear waste transmutation[20,21] and nondestructive inspection of contrabands[22]. Furthermore, tunable attosecond γ-ray pulses will allow probing ultrafast nuclear dynamics.[23]

In the QED regime, γ-photon emission and positron generation in an efficient way have attracted a lot of attention since the topic itself is of great interest and significance.[8,24-30] In this regime, nonlinear Compton scattering (NCS)[31,32] and multi-photon Breit-Wheeler (BW)[33,34] can play an important role to intense γ-ray emission and dense positron production. Note that high-energy γ photons are necessary to trigger the multi-photon BW process. Such photons can still result from radiation loss of relativistic electrons with bremsstrahlung radiation[35-37] and synchrotron/betatron radiation.[38-40] In order to realize effectively γ-photon production and/or positron creation, novel target geometries included wire target[41-45], cone target[46-47], channel target[48-49], sandwiched target[50] and multi-layer target[51-52] have been investigated extensively.

With the development of Nano-Micro array (NMA) fabrication[53], laser interactions with NMAs have been proposed to produce brilliant γ-rays and high-yield positrons with appealing conversion efficiency. When employing the NMAs, the laser can penetrate into them and interact with their side walls within the skin depth. As a result of such volumetric heating, the laser absorption of NMAs is much higher, in contrast to planar targets where the laser pulses only heat the surface of the targets. The solid substrates can play a role of plasma mirror, leading to laser reflection. Kulcsár et al.[53] proposed for the first time intense laser-driven nanowire array on a substrate to produce the brightest X-ray pulses. Purvis et al.[54] investigated the irradiation of nanowire arrays for exploring high-energy-density (HED) physics. Jiang et al.[55] reported on the first successful proof-of-principle experiment to produce and control laser-driven electrons with nanowire array. Curtis et al.[56] demonstrated that the nanowire-array target irradiated by laser beam can dramatically influence the efficiency of conversion of laser energy into energy of hot dense plasma. Rubovic et al.[57] studied that nanowire arrays irradiated by intense laser pulse can induce D-D fusion reactions. More recently, laser-wire interactions have been employed to generate bright synchrotron radiation[58], and attosecond bunches of γ flashes and positrons[59]. The above simulations are limited to the two-dimensional geometry and the conversion efficiencies from laser to particles are unfavorable since the array geometry was not optimized. The interaction of QED-strong laser fields with micro-sized array has not been explored yet.

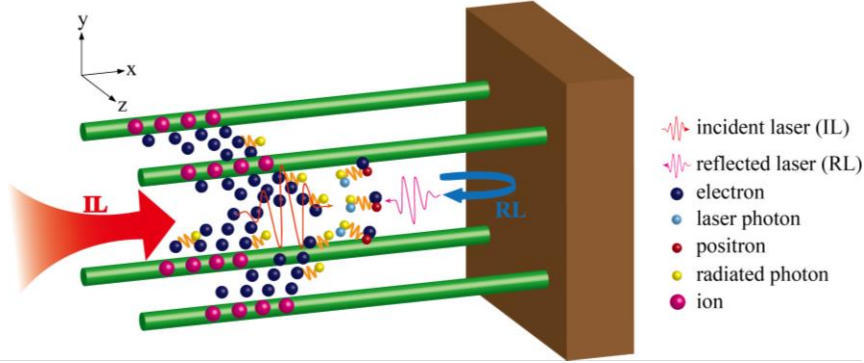

**Fig. 1** The schematic diagram of an ultraintense laser illuminated NMA with substrate incorporated. The incident laser (IL) shoots the NMA at normal incidence to the substrate, which is designed to reflect strongly the incident laser and then to excite fully QED effects.

In this paper, we present an all-optical scheme to produce brilliant attosecond γ-ray flashes and high-yield positron bunches with high conversion efficiency. This is achieved by using an ultraintense laser incident upon a NMA incorporated with a thin substrate, as schematically illustrated in Fig. 1. When the ultra-intense laser pulse is injected normally into the NMA, a large number of electrons are periodically dragged out of the wire surface, forming a train of dense attosecond electron bunches at ultra-relativistic energy. These electrons quiver violently in the strong laser field and emit synchrotron-like radiations. In the following stage, they proceed to collide with intense laser pulse reflected from the substrate and QED effects can be fully excited, which results in generation of very bright γ-ray pulses and high-yield positron bunches. Compared to the laser-wire collision[44], laser traveling inside the NMA is able to enhance the effect of electron heating. With the help of the substrate that acts as a plasma mirror and then realizes colliding geometry, the subsequent γ-ray emission and positron production can be enhanced significantly.

## 2. MODELLING AND RESULTS

To model the electron dynamics and the possible QED effects in the NMAs, full three-dimensional particle-in-cell (3D-PIC) simulations were performed with the QED-PIC code EPOCH[60, 61], where we consider the quantum-modified radiation. The quantum-corrected photon emission and electron-positron pair production via multi-photon BW process have been incorporated successfully, which make behavior of particles modified greatly. The simulation box has a size of 30 μm ($x$) × 20 μm ($y$) × 20 μm ($z$), which is sampled by cells of 1500 × 400 × 400 with 125 pseudo-electrons, eight pseudo-carbons and four pseudo-aluminums in each cell. The Gaussian laser pulse propagates along the $x$-direction with $y$-direction polarization (see Fig. 1), wavelength $\lambda_0 = 1$ μm (laser period $T_0 = 2\pi/\omega_0 =$ 3.33 fs), peak power ~50 PW, duration 30 fs in full width at half maximum (FWHM). With an initial focal spot radius $\sigma_{ini}$ = 3 µm and an amplitude $a_{ini}$ = 540 normalized by $m_e\omega_0 c/e$, the laser pulse is located at 1 Rayleigh length (14.25 µm) ahead of the focusing plane. The focal spot radius at the focusing plane is expected to be $\sigma_0$ = 2.13 µm with $a_0$ = 760 (the corresponding intensity $I_0 = 8 \times 10^{23}$ W/cm$^2$) in the vacuum. Four carbon microwires are taken with length $l$ = 12 µm (in the x direction), radius $r_0$ = 0.5 µm and spacing $d$ = 3.5 µm. These wires have a longitudinal location of $3 < x < l + 3$ with wire front end at $x$ = 3 µm. Note that four isolated wires were used to mimic the NMA, in order to reduce the computational requirements. The wires have an initial density of $n_e = 300n_c$ (where $n_c = 1.1 \times 10^{21}$ cm$^{-3}$ is the critical plasma density) and are attached to a 4-µm-thick aluminum foil (substrate) of $n_{Al} = 25n_c$ density. Such a foil is adopted as a plasma mirror to reflect the laser pulse. Laser pulse is reflected at $x$ = 15 µm (corresponding to time instant $t = 15T_0$). In these simulations, periodic boundary condition was employed for the laser field in the transverse directions and absorbing boundary condition for the particles.

## 2.1 EFFICIENT ELECTRON BUNCH ACCELERATION

In the extreme laser field, radiation-reaction effect must be considered. When laser intensity reaches $I\lambda_0^{4/3} = 10^{23}$ W/cm$^2$ µm$^{4/3}$, the damping force the electrons experience becomes significant enough to compensate for the Lorenz force. The equation of electron's movement after the radiation-reaction correction is as follows[62]

$$\frac{d\mathbf{p}}{dt} = -e(\mathbf{E} + \boldsymbol{\beta} \times \mathbf{B}) - (2e^4/3m_e^2 c^4)\gamma_e^2\boldsymbol{\beta}\{(\mathbf{E} + \boldsymbol{\beta} \times \mathbf{B})^2 - (\mathbf{E} \cdot \boldsymbol{\beta})^2\}, \qquad (1)$$

where $e$ is the charge unit, $m_e$ is the electron mass, and $\boldsymbol{\beta}$ is the normalized electron velocity by the light speed in vacuum $c$, and $\mathbf{E}$ and $\mathbf{B}$ are the magnetic and electric fields, respectively. As the laser pulse is incident on the NMA, electrons can be largely extracted from the wires into the inter-wire voids because geometrically the wires occupy such a small fraction of the interaction volume, as shown in Fig. 2(a). These electrons are bunched on two opposite wires (along y-direction) and on the same polarization plane, and are separated by half of the laser wavelength due to the oscillatory nature of the laser field. Due to the electron motion, a strong sheath field $E_y$ is induced around the wire surface, as shown in Fig. 2(b). In the transverse direction, electrons can only be accelerated within half of laser period and a maximal length for electron transverse acceleration is estimated to be $l_y \approx \lambda_0/2$.[63] Then the sheath field generated by charge separation is roughly given as $E_y = en_e^w l_y/\varepsilon_0 \sim 200$ TV/m, considering the electron bunches have an approximate average density of $n_e^w = 20n_c$. Here $\varepsilon_0$ is the permittivity of vacuum. This shows reasonable agreement with the simulation results of 192 TV/m in Fig. 2(b).

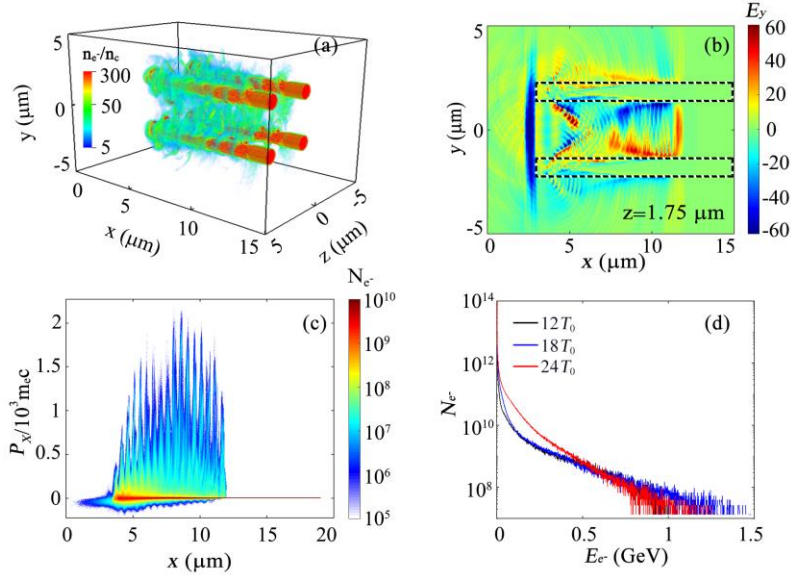

**Fig. 2** (a) Electron density distribution $n_{e^-}$, (b) surface quasi-static electric field $E_y$, and (c) phase-space distribution of electron bunches $(x, p_x)$ at $t = 12T_0$. (d) Temporal evolution of electron energy spectrum. Here, the density and the field strength are normalized by $n_c$ and $E_0 = m_e \omega_0 c / e$, respectively. The electromagnetic (EM) field is averaged over the laser period.

In our geometry, a large number of electrons are pulled out from the skin layer of the wires with a large transverse momentum. Under the ultra-relativistic conditions, the magnetic field force is as strong as the electric field force and the $J \times B$ force turns the transverse motion of electrons into the longitudinal direction. These electrons can further be accelerated to higher energy by the ponderomotive force. They finally exhibit a series of sawtooth-like structures with a large transverse momentum [see Fig. 2(c)], which is similar to the simulation result shown in [44]. This structure indicates that the duration $\tau_e$ of the electron bunches in time enters into the attosecond domain, i. e., $\tau_e = 400$ as at $t = 12T_0$. The cut-off energy of electrons accelerated approaches 1.5 GeV [see Fig. 2(d)], which is visibly higher than 1.0 GeV obtained in the flat target.[64] The total charge of electrons reaches the level of 200 nC and the peak density in each spike is up to $100n_c$. This is one order of magnitude higher than in the gas scheme.[28] Such bunches, with high number-densities and large total charge, would have a beneficial impact in subsequent generation of γ photons. Despite the incident laser pulse does not illuminate directly the wire target, the following electron heating is still quite efficient. The results show that approximately 50% of laser energy is transferred successfully into the kinetic energy of electrons accelerated and γ photons generated. At $t = 24T_0$, the cut-off energy of electron bunches decreases slightly due to the effect of strong radiation reaction force. The more energy the electrons have, the stronger radiation reaction force they are likely to experience, which cuts down the high-energy tail of the electron's spectral distribution.

## 2.2 BRILLIANT ATTOSECOND γ-RAY FLASH EMISSION

In the QED regime, the photon emission probability is characterized by the quantum invariant $\eta_e \approx \gamma_e/E_s |\mathbf{E}_\perp + \mathbf{v} \times \mathbf{B}|$[34]. Here, $\gamma_e$ is the Lorentz factor of the electrons, $\mathbf{E}_\perp$ is the component of the electric field perpendicular to the electron velocity $\mathbf{v}$, $E_s = m_e^2 c^3/e\hbar = 1.3 \times 10^{18}\ V/m$ is the critical field of QED[65] and $\hbar$ is reduced Planck constant. It can be simplified as $\eta_e \approx E/E_s (\gamma_e - p_x/m_e c)$ in the case of an EM plane wave propagating along the x-direction. As the relativistic electrons co-propagate through the IL (denoted as stage I), the $\eta_e$ is minimized to be $(2\gamma_e)^{-1}(E/E_s)$. This is due to the fact that the term $\mathbf{v} \times \mathbf{B}$ can almost cancel out the electric field $\mathbf{E}_\perp$ when referring to the frame of reference boosted to the laser pulse. In contrast, as these electrons travel through the RL pulse (denoted as stage II), the terms $\mathbf{E}_\perp$ and $\mathbf{v} \times \mathbf{B}$ becomes additive, which leads to $\eta_e \approx 2\gamma_e(E/E_s)$. The probability of photon emission by an electron (positron) with energy $\epsilon$ can be expressed as[66,67]

$$W_{rad} \approx 0.46 \frac{\alpha m_e^2 c^4}{\hbar \epsilon} \eta_e, \qquad \eta_e \ll 1, \qquad (2)$$

$$W_{rad} \approx 1.46 \frac{\alpha m_e^2 c^4}{\hbar \epsilon} \eta_e^{2/3}, \qquad \eta_e \gg 1. \qquad (3)$$

Here $\alpha = e^2/\hbar c$ is the fine structure constant. In the limit of $\eta_e \ll 1$, the photon emission can be treated classically, in which the linear Compton scattering process is playing a key role. In the limit of $\eta_e \gg 1$, the NCS process would become dominant over the linear process. Individual photons are then emitted stochastically, manifesting quantum nature of photon emission. The emission probability becomes nonlinearly dependent on electron energy and the laser field strength (see Eq. 3). It is suggested that $\eta_e \geq 0.1$ is essential to induce significant photon emission[50, 60].

Fig. 3 shows the photon energy spectrum and yield at two independent stages for photon emissions. In the stage I, high-energy electrons moving along the direction of laser propagation perform synchrotron-like motion with curvature radius $\rho \approx \lambda_0 \gamma_e/2\pi a_0$ and then emit broadband radiation with cut-off frequency $\omega_c = 3\gamma_e^3 c \rho^{-1}$.[68] At $t = 15T_0$, given by the average energy of electron in the void $\gamma_e \sim 400$, we estimated such synchrotron-like radiations have a maximum energy of ~320 MeV. This is in agreement with the simulation result shown in Fig. 3(a). In the stage II, high-energy electrons counter-propagate through the RL pulse, and the NCS process is initiated successfully. Compared to the photon emission in the stage I, both the photon yield $N_\gamma$ and the laser-to-photon conversion efficiency $\rho_\gamma$ increase rapidly, as seen in Fig. 3(b). After $t = 30T_0$, a portion of photons at the front end start to run out of the right boundary of the simulation box, causing the slight decrease in the photon yield and the conversion efficiency.

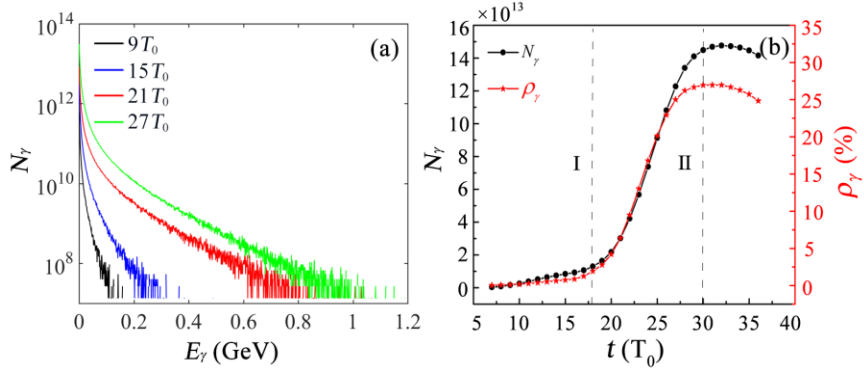

**Fig. 3** Evolution of photon energy spectrum (a), and photon yield and laser-to-photon conversion efficiency (b). The symbols I and II indicate two independent stages for γ-photon emission.

When counting the γ photons with energy above 1.0 MeV, the $N_\gamma$ reaches $1.48 \times 10^{14}$ at $t = 30T_0$. The $\rho_\gamma$ is peaked at ~ 27% accordingly. During the NCS process, the average $\overline{\eta_e}$ can be written as $2(\hbar\omega_0/m_e c^2)\bar{a}_f^2 \sim 0.2$, where $\bar{a}_f^2$ is the amplitude of mean electric fields.[69] Thus, the characteristic photon energy is approximated theoretically by[4, 70]

$$\overline{\hbar\omega_\gamma} \approx 0.44\overline{\eta_e}\bar{\gamma}_e m_e c^2. \tag{4}$$

According to Eq. (4), one can obtain $\overline{\hbar\omega_\gamma}$ ~ 20 MeV at $t = 27T_0$, which is in accordance with the simulation result, 16 MeV.

Temporal evolution of spatial patterns of emitted photons is shown in Fig. 4. Before $t = 21T_0$, photon generation is mainly attributed to transversely oscillating electron synchrotron emission. These photons are produced on both sides of the *x-y* plane due to the initial position of four isolated wires. While most of photons have visibly transverse dispersion; a small portion of photons can propagate stably along the laser axis, which is more clearly shown in Fig. 5(a). At $t = 15T_0$, the photon density distributed on the *z-x* plane is shown in Fig. 4(b). It is obvious that these γ-ray spikes have a similar structure to the attosecond electron bunches. The average duration of these spikes is about 440 as. Note that such spike structure exists visibly in the stage of synchrotron emission but becomes deterioration after $t = 21T_0$. During the stage II, the NCS process dominates the photon emissions and the resultant high-energy photons are in conjunction with the synchrotron-like radiations. Such conjunction finally results in a larger and brighter photon bunch at the center of *x-y* plane, as presented in Figs. 4(c) and (d). Since a large number of photons generated instantly are resided in a very small volume, extremely high photon-densities arise.

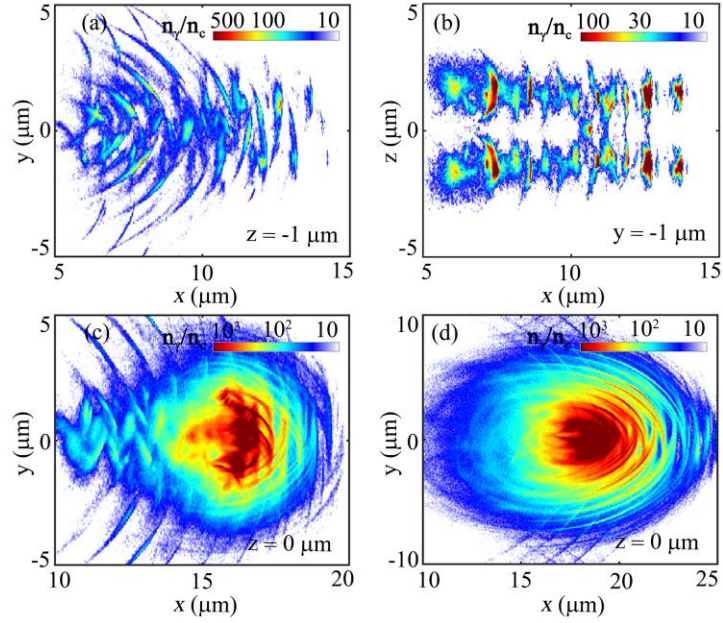

**Fig. 4** Density maps of the emitted photons (contour profile) (a) on the plane of $z = -1$ μm at $t = 15T_0$. (b) Density map of the emitted photons (contour profile) on the plane of $y = -1$ μm at $t = 15T_0$. Density maps of the emitted photons (contour profile) (c) on the plane of $z = 0$ μm at $t = 21T_0$, and (d) on the plane of $z = 0$ μm at $t = 27T_0$.

In the early stage, the emitted photons exhibit the angular distribution featuring two forward lobes at approximately ± 30° [see Fig. 5(a)]. These photons have also a sharp peak angle with respect to the laser axis. This is resulted from high-energy electrons wiggling along the wire surfaces, which generate highly collimated γ-rays.[41] The laser pulse irradiates the flat surface at $t = 15T_0$, hole boring and reflection occur. This is followed by the enhanced photon emission from the NCS process. The relative photon intensity becomes gradually weak in the lobe's direction, and the emitted photons have a collimation angle of ~30°, as shown in Fig. 5(a). In order to calculate photon peak brilliance, we diagnosed photon yield within unit solid angle and unit energy range, i.e. $dI^2/d\Omega dE$. For example, ~$10^{10}$ γ photons were obtained within $10^4$ mrad$^2$ and 15−16 MeV at $t = 27T_0$. The photon source size is 1.5 × 2.9 × 2.2 μm$^3$. With these parameters, the peak brilliance is obtained to be ~$10^{24}$ photons s$^{-1}$ mm$^{-2}$ mrad$^{-2}$ per 0.1% BW @ 15 MeV. Fig. 5(b) shows the peak brilliance as a function of photon energy. It decreases rapidly with increase in photon energy, which is in good agreement with the energy spectra [see Fig. 3(a)]. For different photon energies, the peak brilliances at $t = 27T_0$ are boosted by a factor of 2.5−3.0 compared to the values at $t = 21T_0$. This is caused by a rapid increase in the photon yield [see Fig. 3(b)]. We should note that the brilliance is still above $10^{23}$ with the photon energy of 100 MeV.

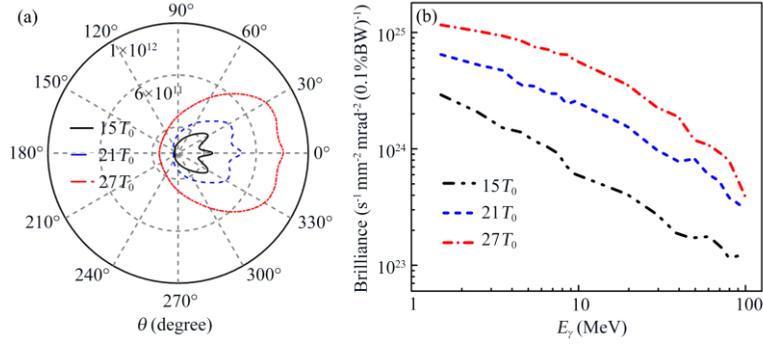

**Fig. 5** (a) Photon intensity patterns on the *x-y* plane and (b) photon peak brilliance depending on photon energy. Here $\theta = \arctan(p_y/p_x)$ represents the polar angle of the emitted photons. For clarity, the photon intensities displayed at $t = 15T_0$ and $21T_0$ are scaled up by a factor of 4 and 2, respectively.

*2.3 HIGH-YIELD POSITRON BUNCH PRODUCTION*

During the stage II, the brilliant γ-ray bursts generated in turn collide with the RL laser, producing a large amount of positrons via the multi-photon BW process. The probability of positron production $W_{pair}$ is determined by the quantum invariant $\chi_\gamma = (1/E_s)\sqrt{(\hbar\omega_\gamma \mathbf{E}/m_e c^2 + \mathbf{k}_\gamma \times \mathbf{B}/m_e c)^2 - (\mathbf{k}_\gamma \cdot \mathbf{E}/m_e c)^2}$, where $\mathbf{k}_\gamma$ is unit vector of the emitted photons[71, 72]. In an EM wave propagating along the *x*-direction, it can be simplified to $\chi_\gamma = \left(\frac{E}{E_s}\right)(\hbar\omega_\gamma - k_x)/mc^2$. Since in our case the colliding geometry is realized, the parameter $\chi_\gamma$ can further be maximized to $2\left(\frac{E}{E_s}\right)(\hbar\omega_\gamma)/mc^2$. This is likely to result in copious electron-positron pair production when the laser pulses have peak intensity exceeding $10^{23}$ W/cm². Fig. 6 show the positron spatial patterns at two instants $t = 24T_0$ and $27T_0$. The positron bunches have maximal density exceeding $10^{22}$ cm⁻³, although they are dispersed slightly in space at a later stage. At $t = 26T_0$, the positrons have decreased energy due to photon emission similar to the electrons in the same laser field. Despite producing at relatively low energies, the positron can be post-accelerated to ultra-high energy up to ~2 GeV, as shown in Fig. 6(c). Meanwhile, they oscillate like electrons in the reflected laser fields and radiate γ photons so that the cut-off energy of positrons reduces visibly from $t = 24T_0$ onwards. The positron yield $N_{e^+}$ reaches peak value of $1.48 \times 10^{11}$ and the laser-to-positron conversion efficiency $\rho_{e^+}$ is about 0.7% [see Fig. 6(d)], which is one order of magnitude higher than in the laser-wire scenarios[44] and in the BH experiments[73]. The positron average energy $\bar{\varepsilon}_{e^+}$ is obtained to be ~360 MeV. Considering positron bunches are concentrated in a cubic volume of ~10 μm³, it generates an energy density as high as $8.5 \times 10^{17}$ J/m³.

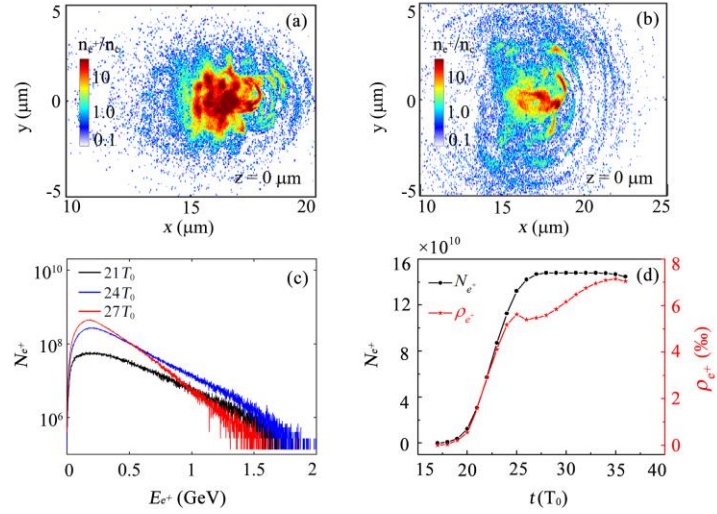

**Fig. 6** Density maps of the created positrons (contour profile) at (a) $t = 24T_0$ and (b) $t = 27T_0$. Evolutions of positron energy spectrum (c) and of positron yield as well as laser-to-positron conversion efficiency (d).

## 3. INFLUENCE OF LASER AND MICROWIRE ARRAY PARAMETERS

A series of 3D-PIC simulations have been performed to study the effect of laser and array parameters on γ-photon and positron production. In the following we will elaborate the production of γ-photons and positrons by changing the laser intensity and the laser-focusing plane, by varying the substrate thickness and by adjusting the wire length, the wire spacing and the wire radius.

*3.1 Changing the laser intensity and the laser-focusing plane*

We first examine the influence of the focused laser intensity (see Fig. 7). As the laser amplitude increases, the $N_\gamma$ and $\rho_\gamma$ increase quickly and at $I_0 = 10^{24}$ W/cm$^2$ the maximum value of $N_\gamma$ and $\rho_\gamma$ reaches $1.8 \times 10^{14}$ and 31%, respectively. The $N_\gamma$ and $\rho_\gamma$ can be approximately scaled as $I_0^{1.5}$ and $I_0^{0.75}$, respectively. The scaling of the conversion efficiency $\rho_\gamma$ can be explained by the following reasons. At $I_0 = 10^{22}$ W/cm$^2$, the parameter $\eta_e$ governing the emission process is proportional to the laser intensity $I_0$.[74] However, as the laser intensity is higher than $10^{23}$ W/cm$^2$, the scaling for the average $\overline{\eta_e}$ is strongly modified to $\overline{\eta_e} \propto I_0^{3/4}$.[8] This leads to the occurrence of conversion efficiency scaling $\rho_\gamma \propto I_0^{3/4}$. It is also found that the $N_{e^+}$ can be scaled as $I_0^{3.5}$ while the $\rho_{e^+}$ follows a similar super-scaling of $I_0^{2.5}$. These scalings indicate that the positron yield has a faster increase compared to the γ photon. In fact, the energy conversion efficiency $\rho_{e^+}$ can be obtained by the total positron energy (the product of positron yield $N_{e^+}$ and the positron average energy $\bar{\varepsilon}_{e^+}$) divided by laser pulse energy $E_L$, i.e. $\rho_{e^+} = N_{e^+} \bar{\varepsilon}_{e^+} / E_L$. The simulation shows that the growth of $N_{e^+}$ is followed by $N_{e^+} \propto I_0^{3.5}$, whereas the value of $\bar{\varepsilon}_{e^+}$ keeps almost constant with the increased laser intensity. The growth expression $\rho_{e^+} \propto I_0^{2.5}$ occurs consequently. Here we should note that it is very challenging to derive theoretically the dependence of $N_{e^+}$ on the laser intensity, since the probability $W_{pair}$ is exponentially suppressed in the classical limit $\chi_\gamma \ll 1$, whereas it becomes significant in the limit $\chi_\gamma \gg 1$. [71]

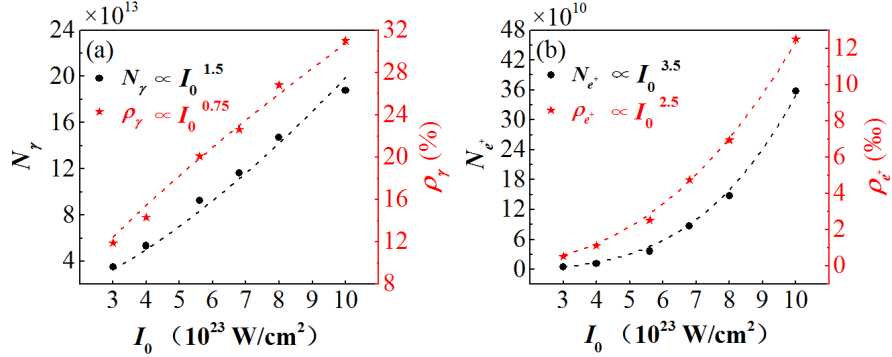

**Fig. 7** Particle yield ($N_\gamma$ and $N_{e^+}$, black circles) and energy conversion efficiency ($\rho_\gamma$ and $\rho_{e^+}$, red stars) as a function of laser intensity at $t = 33T_0$. The wire length is $l = 12$ μm and wire spacing $d = 3.5$ μm.

We now optimize the particle yield and energy conversion efficiency by taking the laser-focusing plane a distance behind the wire front-end. This allows a distance to accelerate and generate well-guided GeV electron beams before reaching the highest laser intensity. The simulation results are shown in Fig. 8. The γ-photon and positron production becomes more efficient when the laser-focusing plane is close to the surface of the substrate. With the laser-focusing position at $x = 14.25$ μm, the $\rho_\gamma$ ($\rho_{e^+}$) reaches a high value of ~27% (0.70%). Such value is enhanced by a factor of ~0.4 (2.0) compared to the one, ~19% (0.25%) obtained at $x = 3.0$ μm, which corresponds to front-end position of the NMA.

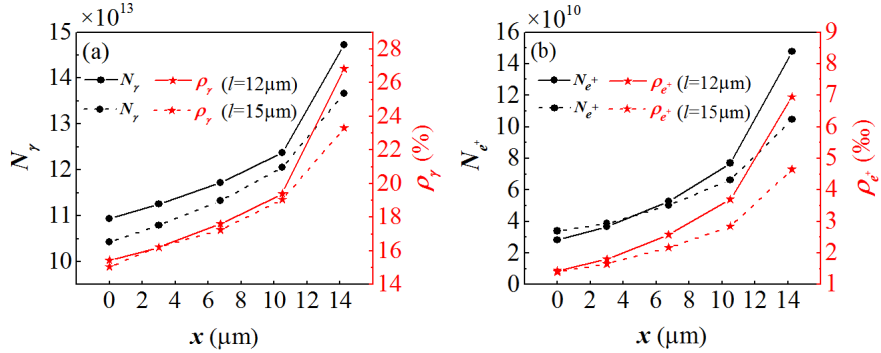

**Fig. 8** Particle yield ($N_\gamma$ and $N_{e^+}$, black circles) and energy conversion efficiency ($\rho_\gamma$ and $\rho_{e^+}$, red stars) as a function of laser-focusing position at $t = 33T_0$. The wire length is $l = 12$ or 15 μm and wire spacing $d = 3.5$ μm.

*3.2 Varying the substrate thickness*

The feasibility of the proposed scheme is further demonstrated by varying the size of NMA.

The substrate thickness is optimized to obtain peak values of both photon and positron yields. When the laser pulse reaches the front surface of the substrate, the laser firstly heats the substrate and laser reflection takes places. However, because the initial density $n_e$ is smaller than the $a_0 n_c$, the target heated becomes relativistically transparent in the later stage. Such effect would affect the laser reflection and then weakens the following QED effects. Our simulation shows that the laser pulse can penetrate a 4-µm-thick substrate when the laser amplitude $a_0 = 760$. The dependence of $N_\gamma$ and $\rho_\gamma$ on the substrate thickness is shown in Fig. 9. When a thinner substrate is employed, the γ-photon and positron production becomes unfavorable due to the target transparency as aforementioned. The $N_\gamma$ and $\rho_\gamma$ first increase rapidly and then get saturated. The $N_{e+}$ and $\rho_{e+}$ have a similar growth trend. However, positron generation has a faster saturation trend compared to γ photon [see Fig. 9(b)] due to a rapid depletion of laser field as discussed later, which is not beneficial for the later multi-photon BW process. This is different to the target transparency induced by two counter-propagating laser pulses, in which positron generation can be enhanced significantly[26]. Our results show that the optimal thickness for the substrate should be 2.5-4.0 µm.

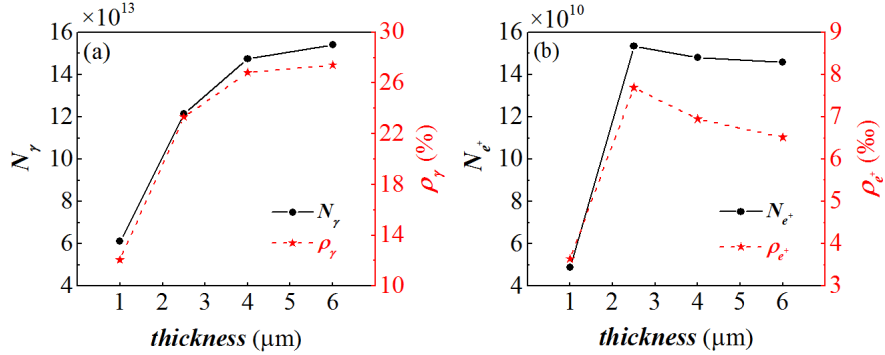

**Fig. 9** Particle yield ($N_\gamma$ and $N_{e+}$, black circles) and energy conversion efficiency ($\rho_\gamma$ and $\rho_{e+}$, red stars) as a function of substrate thickness at $t = 33T_0$. The wire length is $l = 12$ µm and wire spacing $d = 3.5$ µm.

*3.3 Adjusting the wire length, the wire spacing and the wire radius*

The wire length has a significant impact on the particle production, as displayed in Fig. 10. With proper $d = 3.5$ µm, both the $N_\gamma$ and $\rho_\gamma$ first increase and then get decreased when $l > 6$ µm. The $N_{e+}$ has the same variation trend. However, the $\rho_{e+}$ decreases with increasing wire length, as a result of a rapid decrease in the $\bar{\varepsilon}_{e+}$. The underlying physics is that the strong laser field can post-accelerate the positrons more efficiently when the wires become relatively short. In our scenario, an optimal length $l = 12$ µm is obtained for both γ-ray emission and positron production. The occurrence of the optimal length comes from the compromise of electron acceleration and laser reflection. In order to excite fully QED effects, the electrons should be accelerated to ultra-high energy and then interact sufficiently with the RL pulse. However, the electron energy no longer increases with the wire length due to strong radiation-reaction effect. On the other hand, the substrate holding relatively short wires is more favorable to laser reflection, which enhances the following QED effects.

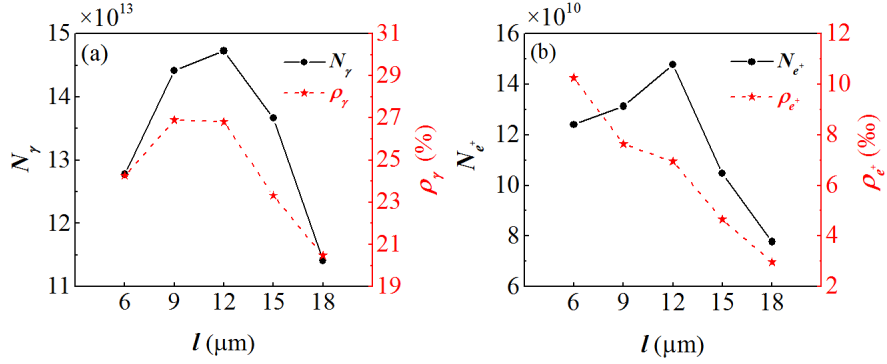

**Fig. 10** Particle yield ($N_\gamma$ and $N_{e^+}$, black circles), energy conversion efficiency ($\rho_\gamma$ and $\rho_{e^+}$, red stars) as a function of wire length $l$ at $t = 33T_0$. The wire spacing is $d = 3.5$ μm and the substrate thickness is 4 μm.

The time evolution of the different components (laser, electrons, ions and photons) is shown in Fig. 11. The laser energy has a rapid depletion after $t = 15T_0$. Due to the substrate heating, the laser energy transfers continuously into electrons and ions. Such rapid depletion is also attributed to tremendous production of γ photons via the NCS process in the stage II. As the laser pulse irradiates the NMA, laser reflection takes place at the front tip of the wires. It further results in small port of laser energy (~10 %) leakages from the simulation boundary during the time frame of $8.5T_0 < t < 11.5T_0$ (see Fig. 11). As a result, the energy balance was obeyed throughout the simulations. Additional simulations show that laser energy depletion occurs in the case of short wires, e.g. $l = 9$ μm.

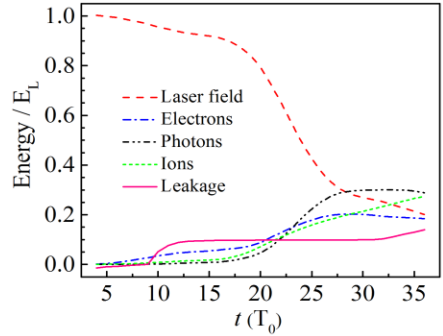

**Fig. 11** Energy of the laser field and particles as a function of time for $a_0 = 760$. $E_L$ denotes the initial energy of the laser pulse. The energy leakage from the simulation boundary is also shown here. The wire parameters are $l = 12$ μm and $d = 3.5$ μm, and the substrate thickness is 4 μm. The data for positron energy is not shown here due to a small portion of energy.

We shall discuss the effect of wire spacing $d$ and wire radius $r$ under the circumstance of wire number $N = 16$ (arranged into double rings with the inner ring having 4 wires and outer ring 12 wires). For $d \geq 3.5$ µm, the outer ring of NMA is immersed in the laser field at exponential weakening intensity, leading to an insignificant QED effects. Simulation results show that there is tiny difference (within 2%) between particle yields (as well as in energy conversion efficiencies) obtained with $N = 4$ (for single ring) and 16 (for double rings). However, as the wire spacing becomes relatively narrow, the single-ring scenario may affect the target heating and the consequent QED effects. For cautious, we take the double-ring scenario to investigate the impact of wire spacing on particle production and the corresponding energy conversion efficiency, as displayed in Fig. 12. As the wire spacing increases, the $N_\gamma$ decreases visibly. However, $\rho_\gamma$ are kept the same value of ~24% when $d \leq 3.5$ µm and then decreases to 18.5% when $d = 5.5$ µm. It is found that both the $N_{e^+}$ and $\rho_{e^+}$ reach their maximum values at $d = 3.5$ µm. Note that the inter-wire voids between adjacent wires are 2.5 µm, which is very close to the focused spot size $\sigma_0 = 2.13$ µm. For $d > 2\sigma_0$, the laser transverse field becomes insufficiently strong to pull out substantial number of wire electrons, hence impeding the excitation of QED effects.

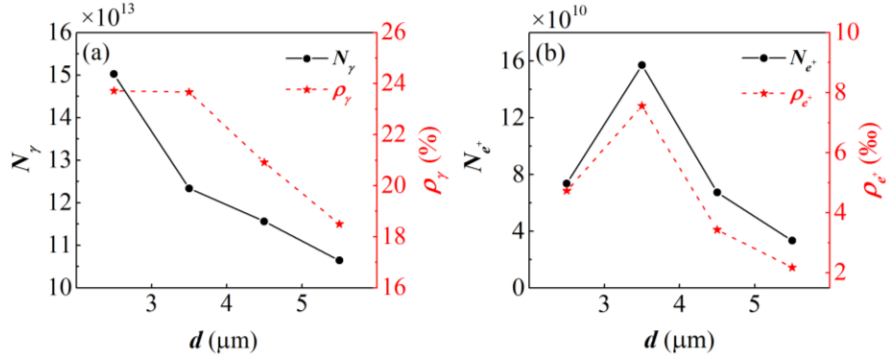

**Fig. 12** Particle yield ($N_\gamma$ and $N_{e^+}$, black circles) and energy conversion efficiency ($\rho_\gamma$ and $\rho_{e^+}$, red stars) as a function of wire spacing $d$ at $t = 33T_0$. The wire length is $l = 12$ µm and the substrate thickness is 2.5 µm, which reduces the huge computational burden for the QED-PIC simulations.

The angular distributions of γ photons are shown in Figs. 13(a)-(b). At $t = 15T_0$, γ photons have better collimation angle (along the direction of laser propagation) when the wire spacing becomes larger owing to the well-collimated electron bunches oscillating in strong laser fields. At a later stage, the collimation effect disappears and the photon bunches have half-opening angle of approximately 30°. Then the wire spacing can hardly affect the angular divergence of the emitted γ photons.

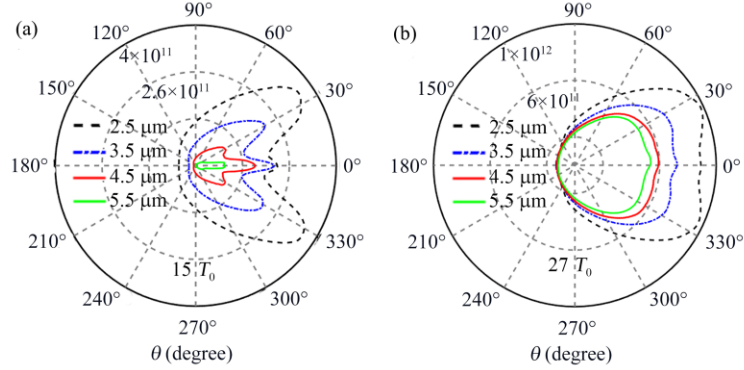

**Fig. 13** Photon intensity patterns on the *x-y* plane at $t = 15T_0$ (a) and $27T_0$ (b), respectively. For clarity, photon intensities displayed for $d = 4.5$ and $5.5$ μm at $t = 15\,T_0$ are scaled up by a factor of 4 and 8, respectively. The wire length is $l = 12$ μm and the substrate thickness is 2.5 μm.

Wire radius *r* is another key parameter to influence the γ-photon emission and positron generation. The electron dynamics in the NMA depend on the static electric field and the laser field amplitude. The effective critical radius can be obtained by $r_{cr}^e = N_L a_d (n_c/n_e) \lambda_0 / 2\pi$, where $N_L$ is the number of laser periods and $a_d = a_0 \exp(-\frac{d^2}{2\sigma_0^2})$ is the laser amplitude experienced by the inner ring of the wires.[75] The meaning of $r_{cr}^e$ is that the laser pulse is capable of complete removal of electrons when $r \leq r_{cr}^e$. In our scenario, we obtain the value $r_{cr}^e \approx 0.57$ μm while $a_d = 200$. The simulation on the wire radius is shown in Fig. 14. The particle yields and the resulting energy conversion efficiencies increase first and then reach peak values at $r = 0.5$ μm, which is in accordance with the value for $r_{cr}^e$.

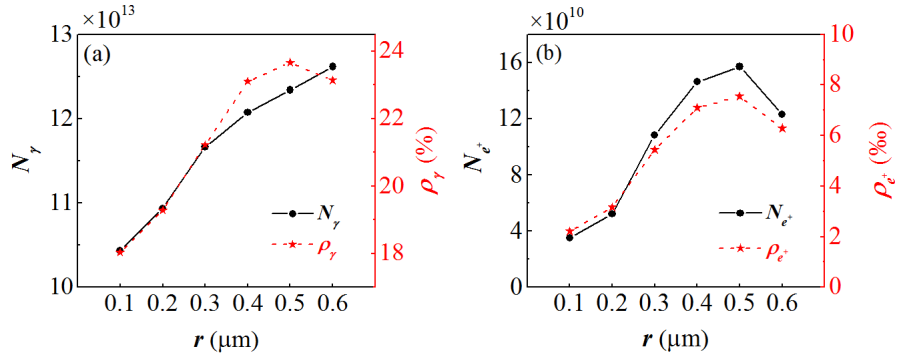

**Fig. 14** Particle yield ($N_\gamma$ and $N_{e^+}$, black circles), energy conversion efficiency ($\rho_\gamma$ and $\rho_{e^+}$, red stars) as a function of wire radius *r* at $t = 33T_0$. The wire spacing and wire length are fixed to $d = 3.5$ μm and $l = 12$ μm, respectively. The substrate thickness is 2.5 μm.

## 4. DISCUSSION

Positron annihilation (into two photons) is an inverse process relative to positron production (via the multi-photon BW process). This annihilation should also exist in the HED plasmas. Here we discuss the effect of positron annihilation on the positron survival in such plasmas. For in-flight $e^+e^-$ annihilation into two photons, the cross section is[76,77]

$$\sigma_{an}(Z,\gamma) = \frac{z\pi r_0^2}{\gamma+1}\left[\frac{\gamma^2+4\gamma+1}{\gamma^2-1}\ln\left(\gamma+\sqrt{\gamma^2-1}\right) - \frac{\gamma+3}{\sqrt{\gamma^2-1}}\right], \quad (7)$$

where $r_0 = 2.82 \times 10^{-13}$ cm is the classical electron radius, $\gamma$ is Lorentz factor of the positron and $Z$ is the atomic number. In our scheme, we have $\bar{\gamma}_{e^+} \sim \sqrt{1+a_0^2}$, recalling that positron momentum is related to the laser field by $a_0 = p/m_e c$ in the present context. In the simulations we have $\bar{\gamma}_{e^+} \sim 700$ and with Eq. (7), $\sigma_{an} \sim 2 \times 10^{-27} Z$ cm$^2$. The annihilation probability within unit plasma length can be expressed as $n_i \sigma_{an}$, where $n_i$ is the number density of ions.

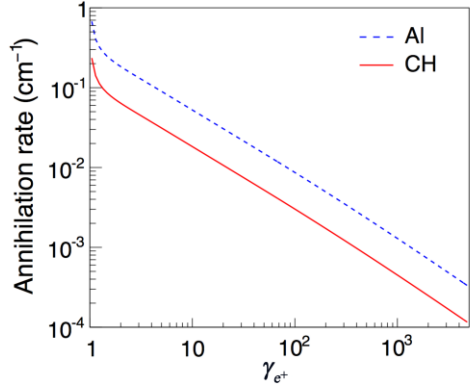

**Fig. 15** Annihilation probability of the created positrons as a function of positron energy.

Fig. 15 shows the annihilation probability for the created positrons as a function of positron energy. One can see that for ultra-relativistic positrons in a low-Z environment, the annihilation rate per centimeter is of the order of $10^{-3}$. This is due to the extremely low annihilation cross section. Hence the positron annihilation effect can be ignored in such plasma conditions. Our scheme is suitable to produce HED positron bunches for the astrophysics study in the laboratory. The simulations show that the maximal positron yield is $1.48 \times 10^{11}$, which result in a peak density of $n_{e^+} \sim 10^{22}$ cm$^{-3}$. Taking $\bar{\gamma}_{e^+} \sim 700$, the plasma frequency $\omega_{e^+} = (n_{e^+} e^2/\bar{\gamma}_{e^+} m_{e^+} \varepsilon_0)^{1/2} \sim 2.1 \times 10^{14}$ s$^{-1}$, where $m_{e^+}$ is the rest mass of the positron. Therefore, the collision-less skin depth of the positron beam is $l_{skin} \approx c/\omega_{e^+}$, about 1.4 μm. This value is slightly smaller than the transverse size $D_{pair} \sim 2.0$ μm [see Fig. 6(b)], indicating that the collective effects of the pair plasma may be also triggered.[78] However, the positron density is still three orders of magnitude less than that required for initiating EM instabilities of astrophysical interests.

We further compare the effectiveness of our scheme with previously proposed regimes in Table 1 by showing the laser amplitude $a_0$, the laser pulse energy $E_L$, the energy conversion efficiencies $\rho_\gamma$ and $\rho_{e^+}$, and the resulting yields $N_\gamma$ and $N_{e^+}$. Compared with the proposed regimes[69,79], our scheme can result in highest laser-to-particle energy conversion efficiency. More specifically, the $\rho_\gamma$ and $\rho_{e^+}$ are two times higher than those obtained in [69], despite the $N_\gamma$ and $N_{e^+}$ have smaller values and the incident laser pulse energy is less. For single laser scenarios, both the laser-to-photon energy conversion efficiency and photon yield obtained in NMA are almost one order of magnitude higher than those in double-layer target[79]. Our simulation results are still comparable to those obtained using single laser with plasma mirror reflection at visibly higher laser intensity ($a_0 = 604$)[80], where the $\rho_\gamma = 21\%$ and $\rho_{e^+} = 0.2\%$ were obtained with 2D-PIC simulations, which may give rise to an overestimation of the energy conversion efficiency.

**Table 1.** The particle yield ($N_\gamma$ and $N_{e^+}$) and the energy conversion efficiency ($\rho_\gamma$ and $\rho_{e^+}$) obtained with close laser amplitudes $a_0$ but different laser pulse energies and interaction schemes.

| Interaction scheme | $a_0$ | $E_L$ | $N_\gamma$ | $\rho_\gamma$ | $N_{e^+}$ | $\rho_{e^+}$ |
|---|---|---|---|---|---|---|
| Two lasers-driven colliding foils[69] | 283 | 1538 J | $7.2 \times 10^{13}$ | 6.0% | $1.6 \times 10^{10}$ | 0.02% |
| Single laser interacting with double-layer target[79] | 468 | 341 J | $1.7 \times 10^{12}$ | 1.4% | $5.6 \times 10^9$ | 0.04% |
| Single laser irradiated NMA (this work) | 468 | 475 J | $3.5 \times 10^{13}$ | 11.9% | $4.5 \times 10^9$ | 0.05% |

To fully apply this scheme and generate highly brilliant γ-rays with energy ranging to 1.0 GeV, proper alignment between the NMA and the driving laser in such experiments is required. 3D-PIC simulation shows that when the laser axis is set to $y = 1$ μm along the $x$-direction, the $N_\gamma$ and $\rho_\gamma$ are decreased by ~5% and ~14%, respectively, despite the generated photons have similar spectrum to that in the case without laser misalignment. Meanwhile, $N_{e^+}$ and $\rho_{e^+}$ are reduced by ~14% and ~35%, respectively. To see how NMA affects the laser-to-particle energy conversion efficiencies, we compared the results to simulations in which the NMA is removed. It is shown that when the laser pulse irradiates the planar substrate, the $N_\gamma$ and $\rho_\gamma$ decrease by ~15% and ~30%, respectively. Meanwhile, both the $N_{e^+}$ and $\rho_{e^+}$ reduce by ~80%. As a result, the NMA is playing a crucial role in the positron production. In addition, we run a set of simulations by varying the laser spot size, while keeping the same laser energy and fixing the laser-focusing plane to the left boundary of the simulation box. The results show that when the focal spot $\sigma_{ini}$ increases from 3 μm to 7 μm, the $N_\gamma$ and $\rho_\gamma$ reduce significantly by ~50% and ~70%, respectively. Moreover, the $N_{e^+}$ and $\rho_{e^+}$ have tremendous drops by a factor of more than 100. It can be readily explained by the reason that laser intensity obeys $I_{ini} \propto 1/\sigma_{ini}^2$ and positron generation follows $\rho_{e^+} \propto I_{ini}^{2.5}$, as shown in Fig. 7(b).

In current study, an ultrathin substrate has been used in order to reduce the computational cost. However, sub-mm-thick substrates were popularly employed in experiments of laser interaction with microstructures at relativistic intensities[55,56]. We should note that the usage of such thick substrate would not affect the applicability of the proposed scheme due to the following two aspects: First, the electron dynamics and the subsequent QED effects will not be changed even if a thicker substrate is employed, with which particle yield may reach saturation values (see Fig. 9). Second, the attenuation of hundreds MeV positrons passing through the thicker substrate can be ignored (see Fig. 15).

## 5. CONCLUSION

In conclusion, we have shown that brilliant attosecond γ-ray flashes and high-yield positron bunches can be efficiently generated from intense laser-irradiated NMA. The electrons from wire surface are rapidly extracted by the laser transverse field and are then accelerated forward to GeV energy by the laser longitudinal field and the Lorentz force. Besides performing synchrotron-like motion, the accelerated electrons counter-propagate through the reflected laser field, leading to a colliding geometry for efficient production of both γ-photons and positrons. Proper laser-focused position and target geometry are found to maximize the γ-photon emission and positron production, which indicates that compared to nanowire array, microwire array is more beneficial for particle generation in the QED-strong laser fields. The scalings of γ-photon and positron generation in laser-wire interactions are further obtained. The simulation shows that at laser intensity of $8 \times 10^{23}$ W/cm$^2$, brilliant γ-ray flashes are emitted with peak brilliance of $\sim 10^{24}$ photons s$^{-1}$ mm$^{-2}$ mrad$^{-2}$ per 0.1% BW @ 15 MeV. The positron yield reaches as high as $1.48 \times 10^{11}$, which results in peak density of $10^{22}$ cm$^{-3}$. It is expected that such γ-ray flashes and positron bunches with unique properties would be beneficial for investigating attosecond nuclear science and HED physics with the coming 10PW lasers, such as the ELI and the SULF.

**ACKNOWLEDGEMENTS**

The EPOCH code was developed as part of the UK EPSRC funded projects EP/G056803/1. All simulations were performed on the National Supercomputer Center of Guangzhou, and the HPC Center of the University of South China. This works is supported by National Natural Science Foundation of China (Grant Nos. 11675075 and 11975214), Natural Science Foundation of Hunan Province, China (Grant No. 2018JJ2315), Youth Talent Project of Hunan Province, China (Grant No. 2018RS3096), and Independent research project of key laboratory of plasma physics, CAEP (Grant No. JCKYS2020212006).

**DATA AVAILABILITY**

The data that supports the findings of this study are available within the article [and its supplementary material].